\documentclass[aip,apl,numerical,reprint,superscriptaddress]{revtex4-1}

\usepackage{graphicx}
\usepackage{latexsym}
\usepackage{amsmath}
\usepackage{amssymb}
\usepackage{amsfonts}
\usepackage{color}

\begin{document}
\bibliographystyle{aipsamp}

\title{Magnetic mapping of defects in type-II superconductors}

\author{S. Mironov}
\affiliation{Moscow Institute of Physics and Technology, 141700 Dolgoprudny, Russia}
\affiliation{University Bordeaux, LOMA UMR-CNRS 5798, F-33405 Talence Cedex, France}
\author{Zh. Devizorova}
\affiliation{Moscow Institute of Physics and Technology, 141700 Dolgoprudny, Russia}
\affiliation{Kotelnikov Institute of Radio-engineering and Electronics RAS, 125009 Moscow, Russia}
\author{A. Clergerie}
\affiliation{University Bordeaux, LOMA UMR-CNRS 5798, F-33405 Talence Cedex, France}
\author{A. Buzdin}
\affiliation{University Bordeaux, LOMA UMR-CNRS 5798, F-33405 Talence Cedex, France}

\date{\today}
\begin{abstract}
Recently it was discovered that the non-uniform Meissner current flowing around the pinning sites in the type-II superconductor induces the unconventional vortex-antivortex pairs with the non-quantized magnetic flux [J.-Y. Ge, J. Gutierrez, V. N. Gladilin, J. T. Devreese, and V. V. Moshchalkov, Nat. Commun. \textbf{6}, 6573 (2015)]. Here we provide the theory of this phenomenon showing that the vortex-like structures originate from the perturbation of the current streamlines by the non-superconducting defect, which results in the generation of the localized magnetic field. The position and the shape of such vortex dipoles are shown to be very sensitive to the defect form. Thus, applying the external magnetic field or current to the superconductor and using, e.g., the high-resolution scanning Hall microscope to measure the stray magnetic field one can plot the map containing the information about the position of the defects and their shape.
\end{abstract}

\maketitle

The control of the Abrikosov vortex pinning is one of the corner-stone problems in the physics of superconducting systems.\cite{Rosenstein,Velez} The reduction of the vortex mobility by a lattice of the pinning centers allows to damp the energy dissipation and substantially increase the critical current,\cite{Anderson,Wang_2016,Xu,Li,Guenon} which is extremely important for application of superconductors in electronics. During the last decades it became possible to create both low- and high-temperature superconductors with various types of natural and artificial defects (holes,\cite{Baert,Cordoba,Wang} grain boundaries,\cite{Song} nanorods,\cite{Llordes,Tarantini,Braccini} imbedded nanoparticles,\cite{Miura} surface grades,\cite{Motta} controllable lattice transformations,\cite{Ray} etc.) which are shown to be effective barriers for the vortex motion. The corresponding enhancement of the critical current appears to be very sensitive to the particular shape and the spatial distribution of the pinning centers. This aims the efforts of both theoreticians and experimentalists at the engineering of the efficient pinning potentials and the extensive study of the magnetic flux behavior simultaneously affected by the pinning sites and the transport current.

Recently the magnetic field induced by the Meissner current flowing around the pinning sites was measured with the high-resolution scanning Hall microscope.\cite{Moschalkov, Moschalkov_2016} It was found that the magnetic contrast near the defects reminds the one for the pair of vortex and anti-vortex. Interestingly, the magnetic flux carried by each pole of such ``vortex dipole'' is not quantized and depends on current. To explain this effect the authors have performed sophisticated numerical simulations based on the time-dependent Ginzburg-Landau equation accounting the non-uniform profile of the Meissner current. Here we provide the simple explanation of this phenomenon within the stationary currents theory. It is based on the fact that each defect being impervious for the Cooper pairs perturbs the streamlines of the superconducting current which gives rise to the well-localized stray magnetic field. Our analysis clearly shows that the formation of the vortex dipoles is not specific to the Meissner state and can be stimulated by the currents of arbitrary nature. We also show that the distribution of the magnetic field along the sample surface contains explicit information about the shape and the position of the pinning centers. This finding forms the ground for simple and direct mapping technique providing the data on the pinning potential in type-II superconductors.

\begin{figure}[b!]
\includegraphics[width=0.35\textwidth]{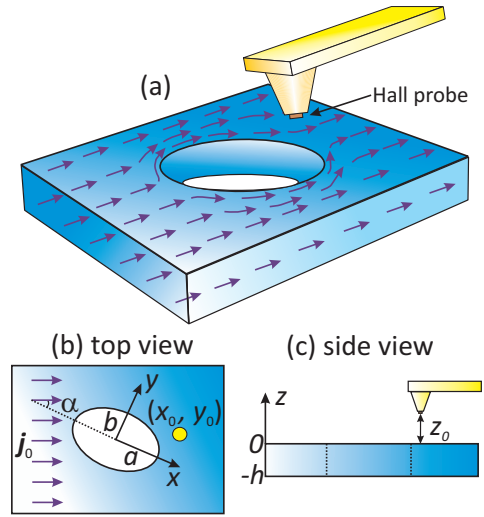}
\caption{(a) The sketch of the superconducting sample with the columnar defect of the elliptic cross-section. The current flowing around the defect produces non-uniform magnetic field which is detected by the tip of the scanning Hall microscope. (b), (c) The top and the side views of the sample, respectively. The external current ${\bf j}_0$ forms the angle $\alpha$ with the $x$-axis containing the semi-axes $a$ of the ellipse. The tip of the Hall microscope is positioned at the point with the coordinates $(x_0,y_0,z_0)$.} \label{Fig_1}
\end{figure}

\begin{figure*}[t!]
\includegraphics[width=0.95\textwidth]{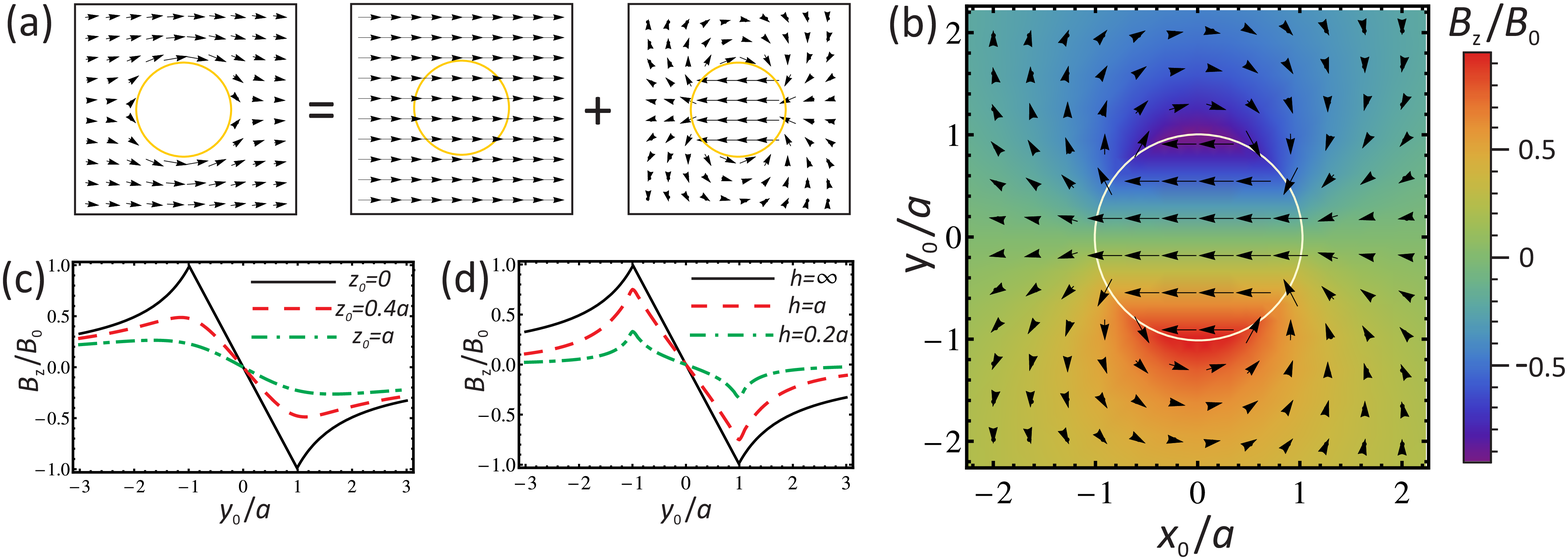}
\caption{(a) The representation of the superconducting current ${\bf j}$ flowing around the defect as the sum of the external uniform current ${\bf j}_0$ which does not induce magnetic field and the deviation $\delta {\bf j}$ which has the form of the vortex-antivortex pair. (b) The dipole-like profile of the perpendicular magnetic field component $B_z$ near the pinning center as a function of coordinates at the sample surface. The white circle of the radius $a$ indicates the boundary of the defect. The external current ${\bf j}_0$ is directed along the $x$-axis. The black arrows show the profile of $\delta {\bf j}$. The superconducting film is assumed to be semi-infinite ($h\to\infty$). (c) The dependence of the magnetic field $B_z$ on the coordinate $y_0$ perpendicular to the current for different distances $z_0$ from the sample surface. (d) The dependence $B_z(y_0)$ at the sample surface for different thicknesses $h$ of the superconducting film. In all panels $B_0=2\pi j_0 a/c$.} \label{Fig_2}
\end{figure*}

Our model system is shown in Fig.~1. The superconducting film of the thickness $h$ contains a non-superconducting columnar defect with the straight line perpendicular to the surface of the film and parallel to the $z$-axis. For simplicity we restrict ourselves to the case when the cross-section of the defect has the form of the ellipse $(x/a)^2+(y/b)^2\leq 1$ with the semi-axes $a$ and $b$ ($a>b$).

Our goal is to calculate the magnetic field outside the film induced by the superconducting current flowing around the defect. Previously, the distribution of the supercurrent affected by the sample boundaries of different shape (for both single- and multiply-connected geometries) was extensively studied on the base of the self-consistent solution of the London and Maxwell equations \cite{Via, Schuster, Gurevich, Friesen}. Typically within such approach one deals with the complicated integral-differential equations which can be solved numerically or with the help of the phenomenological material relations instead of the exact London equation.

Here we significantly simplify the problem assuming that the size of the defect is much smaller than the London penetration depth $\lambda$. This allows to neglect the screening of the stray magnetic field and obtain transparent analytical expressions for the current profiles using the electrostatic analogy,\cite{BuzdinFeinberg1,BuzdinFeinberg2} which has been successfully applied for the description of the interaction between Abrikosov vortices and columnar pinning centers.\cite{BuzDau1,BuzDau2} In this case the current density $\bf{j}$ satisfies the equation ${\rm rot}~{\bf j}=0$ with the boundary condition requiring the absence of the current component perpendicular to the defect surface. We assume that at the distance $r\gg a$ (and, at the same time, $r\ll \lambda$) from the defect center the current is uniform and its density is ${\bf j}_0=j_0\left(\cos\alpha\hat{\bf x}+\sin\alpha\hat{\bf y}\right)$. Then the current distribution ${\bf j}(x,y)$ outside the defect can be represented in the compact complex form \cite{Hydro}
\begin{equation}\label{Current}
j_x-ij_y=j_1-j_2\frac{x+iy}{\sqrt{(x+iy)^2-\left(a^2-b^2\right)}},
\end{equation}
where $j_1=j_0\left(a\cos\alpha+ib\sin\alpha\right)/(a-b)$, $j_2=j_0\left(b\cos\alpha+ia\sin\alpha\right)/(a-b)$, and $i$ is the imaginary unit.

\begin{figure*}[hbt!]
\includegraphics[width=0.95\textwidth]{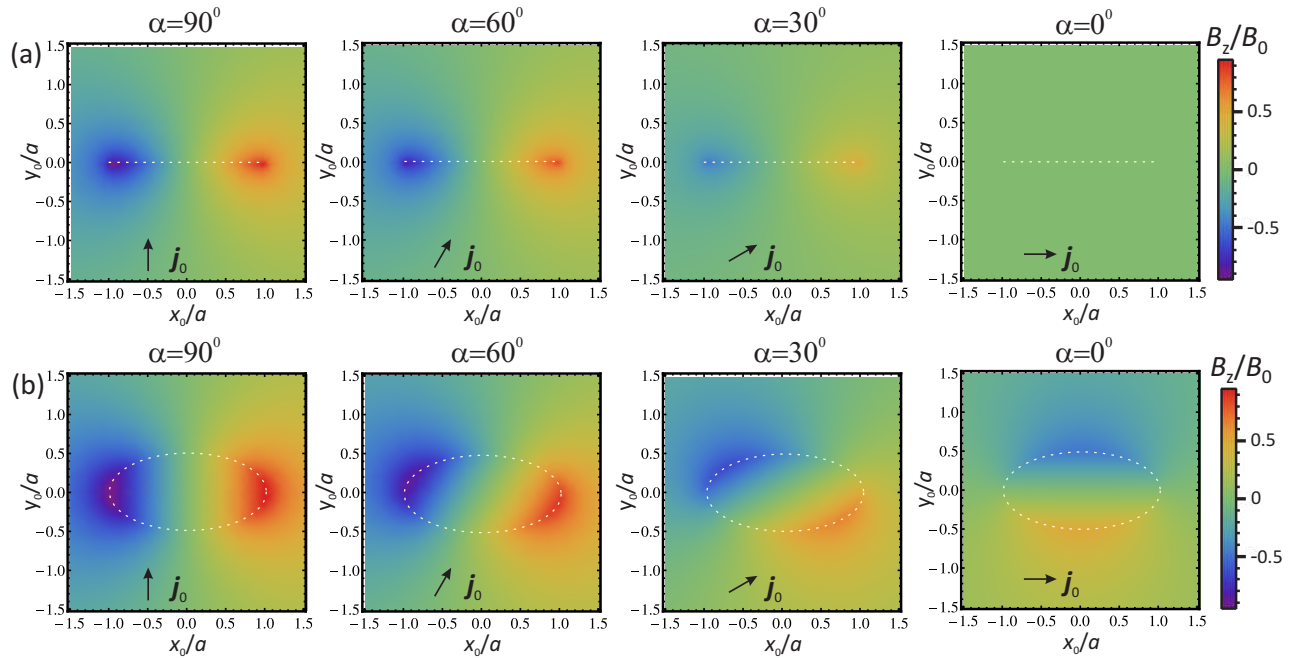}
\caption{The dependence of the magnetic field profile near the pinning center on the shape of the defect and the orientation of the external current. (a) The profiles of $B_z$ at the surface of the semi-infinite superconducting film containing the crack-like defect of the length $a$. Different pictures correspond to different angles $\alpha$ between the external current ${\bf j}_0$ and the straight line containing the defect. (b) The profiles of $B_z$ near the defect of elliptic cross-section. In both panels the white dotted lines indicate the defect boundaries.} \label{Fig_3}
\end{figure*}

The current (\ref{Current}) induces the non-uniform magnetic field ${\bf B}$ localized near the defect. The $z$-component $B_z$ of such field can be directly measured, e.g., by the tip of the scanning Hall microscope. To model such situation we calculate $B_z$ at the point with the coordinates ($x_0$, $y_0$, $z_0$), where $z_0$ is the distance between the tip and the sample surface (see Fig.~\ref{Fig_1}). Note that the  current ${\bf j}$ defined by (\ref{Current}) can be formally represented as the sum of the external uniform current ${\bf j}_0$ which does not induce the magnetic field and the deviation $\delta {\bf j}$ which has the form of the vortex-antivortex pair [see Fig.~\ref{Fig_2}(b)]. Substituting the expression for $\delta {\bf j}$ into the Biot-Savart law and integrating over the sample volume we obtain the analytical expression for the magnetic field component $B_z$. Before analyzing the general situation we will first consider two specific types of defect: the columnar pinning center with the circular cross-section ($b=a$) and the linear crack in the superconductor ($b=0$). To make the analytical results more transparent hereinafter we consider only the case when the sample thickness $h\to\infty$ and the distance $z_0$ between the tip and the sample is negligibly small.

If the cross-section of the defect is circularly symmetric ($b=a$) the profile of $B_z$ depends only on the current direction but not on the orientation of the defect so without the loss of generality we put $\alpha=0$. In this case introducing the polar coordinates $(r,\varphi)$ so that $x_0=r\cos\varphi$, $y_0=r\sin\varphi$ we find:
\begin{equation}\label{Circle_result}
B_z(r,\varphi)=\left\{\begin{array}{l}{-\frac{2\pi j_0}{c} r\sin\varphi~~~~~{\rm for}~~r<a,}\\{-\frac{2\pi j_0}{c}\frac{a^2}{r}\sin\varphi~~~{\rm for}~~r>a.}\end{array}\right.
\end{equation}
This profile of the magnetic field is shown in Fig.~\ref{Fig_2}. Clearly, the magnetic contrast reflects the vortex structure of the current $\delta{\bf j}$ induced by the pinning center. The poles of this vortex dipole are positioned at the defect boundary $r=a$ and correspond to $\varphi=\pm \pi/2$ so that the straight line connecting the poles is perpendicular to the external current ${\bf j}_0$. Moreover, in contrast with the usual Abrikosov vortex each pole contains the magnetic flux which is not quantized and is determined by the external current $j_0$. Thus, our simple model based on the stationary current theory fully explains the formation of the vortex-antivortex pairs recently observed with the scanning Hall microscope.\cite{Moschalkov, Moschalkov_2016}

Note that the finite sample thickness $h$ and the finite distance $z_0$ between the tip and the surface of the superconductor does not lead to any qualitative changes of the $B_z(r,\varphi)$ profiles [see Fig.~\ref{Fig_2}(c) and 2(d)]. Specifically, the decrease of $h$ results in the damping of the magnetic field value and the smothering of its profile. The increase of the distance $z_0$ is also accompanied by the damping of $B_z$, but the magnetic field profile across the defect qualitatively remains the same.

Now we turn to another limiting case $b\ll a$ corresponding to the linear crack in the superconductor. In this case the magnetic field profile strongly depends on the orientation of the defect relative to the direction of the external current. Then the magnetic field profile can be represented in terms of the elliptic coordinates defined as $x_0=a\cosh\mu\cos\theta$, $y_0=a\sinh\mu\sin\theta$:
\begin{equation}\label{Line_result}
B_z(\mu,\theta)=\frac{2\pi j_0a}{c} e^{-\mu}\cos\theta\sin\alpha.
\end{equation}
The corresponding magnetic contrast is shown in Fig.~\ref{Fig_3}(a). Contrary to the case of the circular defect where the position of the vortex and antivortex depend on the external current direction here the poles of the vortex dipole always appear near the ends of the crack independently on the vector ${\bf j}_0$ orientation. At the same time, the maximum value of the magnetic field is proportional to $\sin\alpha$ which allows to extract the information about the crack orientation from the measurements of the magnetic contrast for different ${\bf j}_0$ directions.

Finally, when the pinning center has the elliptic cross-section the expression for the magnetic field strongly depends on whether the tip is positioned above the defect or above the superconducting region. In the first case the component $B_z$ of the magnetic field is
\begin{equation}\label{Gen_result_int}
B_z=\frac{2\pi j_0}{c} \left(x_0\sin\alpha-y_0\cos\alpha\right),
\end{equation}
while in the second case
\begin{equation}\label{Gen_result_ext}
B_z=\frac{2\pi j_0}{c} \sqrt{\frac{a+b}{a-b}}e^{-\mu}\left(a\cos\theta\sin\alpha-b\sin\theta\cos\alpha\right),
\end{equation}
where $\mu$ and $\theta$ are the elliptic coordinates characterizing the position of the microscope tip: $x_0=\sqrt{a^2-b^2}\cosh\mu\cos\theta$, $y_0=\sqrt{a^2-b^2}\sinh\mu\sin\theta$. The magnetic contrast (\ref{Gen_result_int})-(\ref{Gen_result_ext}) for different orientations of the external current with respect to the ellipse is shown in Fig.~\ref{Fig_3}(b).


Interestingly, the defects can provide more favorable conditions for the vortex creation compared to the boundaries of uniform superconductors. Indeed, for the flat surface of the superconductor Abrikosov vortices enter the sample when the current exceeds the depairing current $j_c$. At the same time, the pinning center strongly increases the current density at the certain points of its boundary. In the case of the circular cross-section the current density at the points $\varphi=\pm\pi/2$ is doubled compared to $j_0$. As a consequence, even the current $j_0\sim j_c/2$ should produce the pair of Abrikosov vortex and antivortex. The modification of the current density profile is even more dramatic for the case of elliptic defect. If $\alpha=\pi/2$ the maximal current density $j_m$ monotonically increases with the increase in the ratio $a/b$: for $\varphi=\pm\pi/2$ one finds $j_m=j_0\left(1+a/b\right)$. This local increase in the current density should strongly damp the energy barrier for the entry of the vortex-antivortex pairs. The experimental verification of this prediction can be performed with the scanning Hall microscope which recently allowed to observe the depinning of the Abrikosov vortex from the non-superconducting defect in the presence of the external Meissner current.\cite{Moschalkov_2016}

The sensitivity of the magnetic field profiles to the form of the defects makes it possible to develop the magnetic mapping technique which can provide the direct information about the position of the pinning centers in the superconductor, their shape and orientation. Remarkably, for arbitrary straight line in the $(x_0,y_0)$-plane containing the ellipse center the local maxima of the magnetic field $\left|B_z\right|$ indicate the defect boundaries. This feature is robust against the decreasing the film thickness $h$ or increasing the distance between the film surface and the tip of the Hall microscope $z_0$. Thus, analyzing different cross-sections of the measured magnetic contrast one can find the position of the center of the pinning site (as the center of the line between two poles) and then reconstruct the profile of the defect boundary within the elliptic approximation. The proposed method can become a convenient tool for direct mapping of the columnar pinning sites in the type-II superconductors.

Thus, we demonstrate that the vortex-antivortex magnetic dipoles recently observed with the scanning Hall microscope \cite{Moschalkov, Moschalkov_2016} originate from the stationary current flow around the pinning center. The spatial profile of the non-uniform magnetic field induced by the current contains the explicit information about the shape of the defect. Using the model of the columnar defect with elliptic cross-section we found analytical expressions for the stray magnetic field and analyzed how magnetic contrast measured by the Hall microscope depend on the direction of the external current. Our results provide the platform for the realization of direct and simple magnetic mapping technique, which allows to reconstruct the spatial distribution of the columnar pinning sites inside the particular superconductor and analyze the shape of their cross-section.

This work was supported by the French ANR ``MASH" and NanoSC COST Action MP1201.

\end{document}